\documentclass{amsart}
\usepackage{graphicx}
\usepackage{hyperref}
\usepackage{enumerate}
\usepackage{amsmath}
\usepackage{amssymb}

\newcommand{\NA}{\mathbb N}
\newcommand{\RE}{\mathbb R}

\newcommand{\ve}{\varepsilon}

\newcommand{\la}{\lambda}

\newcommand{\al}{\alpha}
\newcommand{\ome}{\omega}
\newcommand{\lf}{\left}
\newcommand{\ri}{\right}

\DeclareMathOperator{\sech}{sech}

\DeclareMathOperator{\dn}{dn}
\DeclareMathOperator{\cn}{cn}

\begin{document}

\title{Topology induced bifurcations for the NLS on the tadpole graph}

\author{Claudio Cacciapuoti}
\email{claudio.cacciapuoti@uninsubria.it	}
\address{Dipartimento di Scienza e Alta Tecnologia, Universit\`a dell'Insubria, Via Valleggio 11, 22100 Como, Italy, EU}

\author{Domenico Finco}
\email{d.finco@uninettunouniversity.net}
\address{Facolt\`a di Ingegneria, Universit\`a Telematica Internazionale Uninettuno,  Corso Vittorio Emanuele II 39, 00186 Roma, Italy, EU}

\author{Diego Noja}
\email{diego.noja@unimib.it} 
\address{Dipartimento di Matematica e Applicazioni, Universit\`a  di Milano Bicocca,  via R. Cozzi 53, 20125 Milano, Italy, EU}

\begin{abstract}
In this paper we give the complete classification of solitons for a cubic NLS equation on the simplest network with a non-trivial topology: the tadpole graph, i.e. a ring with a half-line attached to it and free boundary conditions at the junction. The model, although simple, exhibits a surprisingly rich behavior and in particular we show that it admits:
1) a denumerable family of continuous branches of embedded solitons bifurcating from linear eigenstates and threshold resonances  of the system; 
2) a continuous branch of edge solitons displaying a pitchfork symmetry breaking bifurcation at the threshold of the continuous spectrum; 
3) a finite family of continuous branches of solitons without linear analogue. 
All the solutions are explicitly constructed in terms of elliptic Jacobian functions. Moreover we show that families of nonlinear bound states of the above kind continue to exist in the presence of a uniform magnetic field orthogonal to the plane of the ring when a well definite flux quantization condition holds true.
Finally we highlight the role of resonances in the linearization as a signature of the occurrence of bifurcations of solitons from the continuous spectrum.
\end{abstract}

\maketitle
\section{Introduction}
The Non-Linear Schr\"odinger Equation (NLS in the following) is one of the better known effective models for several physical systems, from nonlinear optics to BEC. One of its most interesting features is that it admits solitons, which in the focusing case (the one studied here) can be freely translating or pinned at definite sites  when inhomogeneities are present. In the latter case they are usually called {\it standing waves} by analogy with the eigenstates of the linear Schr\"odinger equation, and as in that case they have the form $\Psi(t,x)=e^{-i\omega t}\Phi_{\omega}(x)$. The purely spatial profile $\Phi_{\omega}$ is a nonlinear bound state solving the stationary NLS and associated to the ``frequency" $\omega$. Solitons are a nonlinear phenomenon, and they often bear no apparent relation with the linear structures underlying the complete dynamics. Nonetheless in many cases, their appearance is strongly influenced by the spectrum of the linear dynamics. In the standard NLS on the line or in higher dimension standing waves exist only when the frequency $\omega$ does not fall into the continuous spectrum of the linear part of the equation. The standard family of translating solitons on the line can be considered as bifurcating out of the edge of the continuous spectrum ``in the semi-infinite gap". Pinned solitons at a defect instead are typically the result of a bifurcation from isolated negative eigenvalues \cite{RW}. Standing waves with frequencies in the continuous spectrum of the linear part are not expected because the resonance with the continuous modes should prevent a balance between dispersion and focusing. This is usually the case, nonetheless in the late 90s it was shown, numerically and analytically that in some special models such {\it embedded} solitons exist. Embedded solitons were then discovered in several models, related to higher derivative NLS and KdV, Thirring model, second harmonic generation system and others \cite{YMK99}. Various properties of embedded solitons have been investigated with numerical and partially analytical tools, but a theoretical framework seems to be lacking, as regards their existence, mechanism of generation and stability (see \cite{Y} and reference therein). 

In the present paper we show that a rich soliton phenomenology, where all mechanisms quoted above are in action, arises in the {\it standard} cubic NLS when the topology of the medium sustaining propagation is non trivial. NLS propagation on networks is a growing subject especially due to experimental progresses in confinement of Bose-Einstein condensates in quasi-one-dimensional traps and engineering techniques of condensate wavefunctions to create solitons (see for example \cite{D} and references therein). Until now the problem has been successfully considered at a theoretical level only in simple cases of Y-junctions or H-junctions (see \cite{ACFN4} and reference therein). Notable exceptions are \cite{GSD}, where a mixed linear/nonlinear Schr\"odinger system is considered numerically, showing interesting nonlinearity induced phenomena in particular regarding the appearance of strong resonances with possible applications to soliton propagation in optical fiber networks, and the recent \cite{AST} where a first general analysis of existence and nonexistence of NLS ground states on networks is given. The presently studied model is a cubic NLS on a Tadpole Graph (TG in the following), the simple structure obtained  by attaching a half-line  to a ring. For this model we give a complete classification  of its nonlinear stationary states and bifurcations. The obtained results seem to be typical when an infinite system is coupled to a bounded one and therefore they point to 
a much wider range of application. In our model a denumerable continuous family of embedded solitons bifurcate from linear {\it embedded} eigenstates and a further family of edge solitons bifurcate from the zero energy resonance at the threshold of continuous spectrum; moreover the former families display a secondary bifurcation with spontaneous symmetry breaking exactly at threshold. Finally, other soliton families without linear analogue exist and are described as well. A primary signature of soliton bifurcations seems to be played by threshold resonances, i.e., quasi-eigenstates bounded but not square integrable, of the linearized operators. This rich phenomenology strongly contrasts with the well known case of the standard NLS on the line with or without external potential. On physical grounds this complexity can ultimately be traced back to the coupling between propagation on bounded regions, where periodic behavior is possible, and unbounded regions, where dispersion dominates but focusing nonlinearity allows soliton behavior. 

The model can be further enriched introducing an external physical control parameter given by a magnetic field acting on the head of the TG in the normal direction to its plane. The relevant fact is that the embedded solitons and their bifurcations from the edge survive in the presence of a topological constraint on the flux of the magnetic field. 

\begin{figure}[t]
\includegraphics[width=0.6\textwidth]{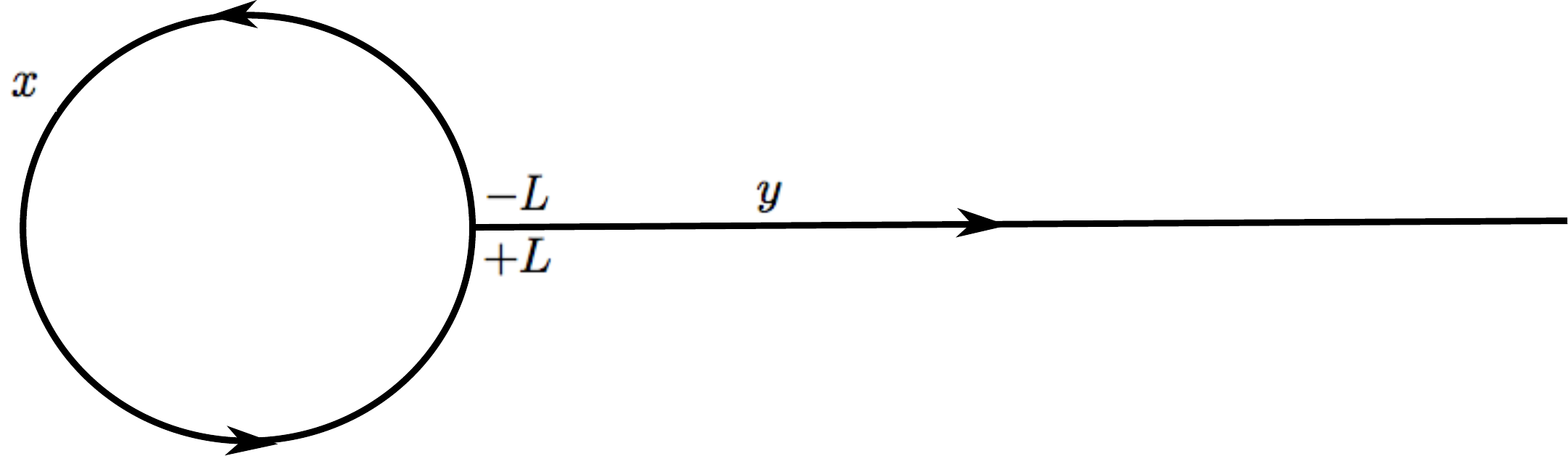}
\caption{\label{f:1}Tadpole graph}
\end{figure}

Now we set up the model. We denote by $x\in[-L,L]$ the coordinate on the ``head'' of the TG and by $y\in[0,\infty)$ the coordinate on the ``tail''. The junction coincides with $x=\pm L$ and $y=0$, see Fig. \ref{f:1}. 
A function $\Psi=(u,\eta)$ on TG has the first component on the head  and the second on the tail.
The scalar product in the $L^2$ space associated to the TG and the corresponding norm are naturally defined as $\langle\Psi_1,\Psi_2\rangle = (u_1,u_2)_{L^2(-L,L)} +  (\eta_1,\eta_2)_{L^2(\RE_+)}$ and $\|\Psi\| = \sqrt{\langle\Psi,\Psi\rangle}$. 

On this structure we consider the focusing NLS for the function $\Psi_t=(u_t,\eta_t)$
\begin{equation}\label{eq0}
i\frac{\partial}{\partial t}\Psi_t=H\Psi_t - |\Psi_t|^2 \Psi_t ,
\end{equation}
where the Hamiltonian $H$ is given by the selfadjoint operator: $H \Psi = (-u'',-\eta'')$
on the domain characterized by the so called free or Kirchhoff boundary conditions at the junction
\begin{align}
D(H) = \Big\{&\Psi=(u,\eta)\in H^2:
u(L)=u(-L)=\eta(0)\label{bc1}\\
&-u'(L)+u'(-L)+\eta'(0) = 0\label{bc2}\Big\}
\end{align}
and $H^2$ is the space of square integrable functions with two square integrable derivatives on each edge. The Hamiltonian $H$ describes a typical and simple {\it quantum graph} \cite{BK, KS, E} with conservation of the probability flux at the junction.  The nonlinear term $|\Psi_t|^2\Psi_t$ is understood as a notation for $(|u_t|^2u_t, |\eta_t|^2 \eta_t)$. It is worth noticing that the model can be rewritten as a system  of two NLS coupled by the boundary condition at the junction. The boundary conditions \eqref{bc1} and \eqref{bc2} are sometimes referred to as \emph{Kirchhoff} boundary conditions and are the most commonly used in applications.  The  spectrum of $H$ is given by $\sigma(H) =[0,\infty)$.  A remarkable fact about the spectrum  of $H$ is  that  $\la_n=(n\pi/L)^2$, $n\in\NA$, are eigenvalues embedded in the continuous spectrum with corresponding eigenfunctions
\begin{equation}\label{Ups}
\Upsilon_n(x,y) =  \big(\sin(n\pi x/L),0\big) \qquad n\in\NA.
\end{equation}
Notice that eigenfunctions corresponding to embedded eigenvalues are confined in a bounded region (here the ring), a fact that is characteristic of quantum graphs (see \cite{KS}) and cannot occur in the simpler topologies of three or lower dimensional space. A second relevant fact is that $H$ has a threshold resonance, i.e., there exists a function $\Upsilon_{res}$ such that $\Upsilon_{res}\notin L^2$, which satisfies the  conditions \eqref{bc1} and  \eqref{bc2}  and such that $H\Upsilon_{res} = 0 $ in distributional sense. 
The resonance function is the constant function
$
\Upsilon_{res} = (1,1)
$.

\section{Nonlinear stationary states}
Standing waves for Eq. \eqref{eq0} are solutions of the form  $\Psi_t=e^{-i\omega t}\Phi_{\ome}$, where purely spatial $\Phi_{\ome}$ satisfies the stationary NLS equation 
\begin{equation}\label{eq}
H\Phi - |\Phi|^2 \Phi =  \ome \Phi \qquad \ome \in\RE\,,\;\Phi\in D(H).
\end{equation}
To compute the stationary solutions it will be convenient to rewrite the equation above componentwise. Set $\Phi=(u,\eta)$, then  Eq. \eqref{eq} is equivalent to the system 
\begin{equation}\label{sistema}
\left\{\begin{aligned}
-&u'' -|u|^2u = \omega u \quad &&x\in[-L,L]\\
-&\eta'' -|\eta|^2\eta = \omega \eta &&y\in[0,\infty)\\ 
&u(L)=u(-L)=\eta(0) &&\\ 
-&u'(L) + u'(-L) +\eta'(0)=0 &&
\end{aligned}\right. 
\end{equation}
with $\ome\in\RE$.  To find the stationary states we find first the general solutions of each equation in  \eqref{sistema} for $u$ and $\eta$. Then we adjust the free parameters  to match the boundary conditions. 

We start with the general  solution on the tail.  A non vanishing, and decaying at infinity solution of  
$
-\eta'' -  \eta^3 = \ome \eta 
$
exists only for $\ome < 0$; such a solution, {\it up to translation and phase multiplication}, is  $\phi(y) = \sqrt{2|\ome|}\sech(\sqrt{|\ome|} y), \quad \ome<0$.

Next we find the general solution of $-u'' -  u^3 = \ome u$ on the head of the TG.  
Since the domain is bounded, a qualitative phase plane analysis gives immediately that solutions can exist for every real $\omega$ and that one should distinguish positive, negative, and vanishing $\omega$. To find their explicit form, one can skip the qualitative analysis and  proceed directly to solve the stationary cubic equation by quadrature method. In this way it can be easily shown that the general solutions are given by  Jacobian elliptic functions (see \cite{Carr} and \cite{SMKJ} for a study of NLS on a ring), and it turns out that it is enough to consider two types of solutions (selected among bounded and independent Jacobian elliptic functions).  The first type exists for $\ome\in\RE$ and, {\it up to translation and phase multiplication}, is given by cnoidal type functions (see \cite{OLBC10})
\[
u_{cn}  (x;k) =  \sqrt{\frac{2\ome k^2}{1-2k^2}} \cn \left( \sqrt{ \frac{\ome}{1-2k^2} }x ;k \right) \qquad \ome\in\RE,
\]
where $k\in(0,1]$ is a free parameter.  The $\cn$ function oscillates between $1$ and $-1$ and therefore the solution $u_{cn}  (x;k)$ has nodes.
These solutions have period $T_{cn}(k) = 4  \sqrt{ (1-2k^2)/\ome }\, K(k)$ where $K$ is the Legendre's complete elliptic integral of the first kind \cite{OLBC10}
\begin{equation}\label{K}
K(\gamma) = \int_0^1 \frac{1}{\sqrt{(1-t^2)(1-\gamma^2t^2)}} dt.
\end{equation}
We remark that for $\omega>0$ (resp. $\omega<0$) $u_{cn}$ and $T_{cn}$ are defined only for $k\in(0,1/\sqrt 2)$ (resp. $k\in(1/\sqrt 2,1)$). 
  
The  second type of periodic solutions exists for $\ome<0$ and is given by dnoidal type functions (see \cite{OLBC10})
\[
u_{dn} (x;\kappa) =  \sqrt{\frac{2|\ome|}{2-\kappa^2}} \dn \left( \sqrt{ \frac{|\ome|}{2-\kappa^2} }x ;\kappa \right) \qquad \ome<0, 
\]
where $\kappa\in[0,1]$ is a free parameter. The $\dn$ function oscillates between 1 and $\sqrt{1-\kappa^2}$ and it is bounded away from 0, and so are the solutions $u_{dn} (x;\kappa)$.  The period of $u_{dn}$ is
$
T_{dn}(\kappa) = 2  \sqrt{ (2-\kappa^2)/ |\ome|}\, K(\kappa)
$.\\

\begin{figure}[h!]
\fbox{\includegraphics[width=0.25\textwidth]{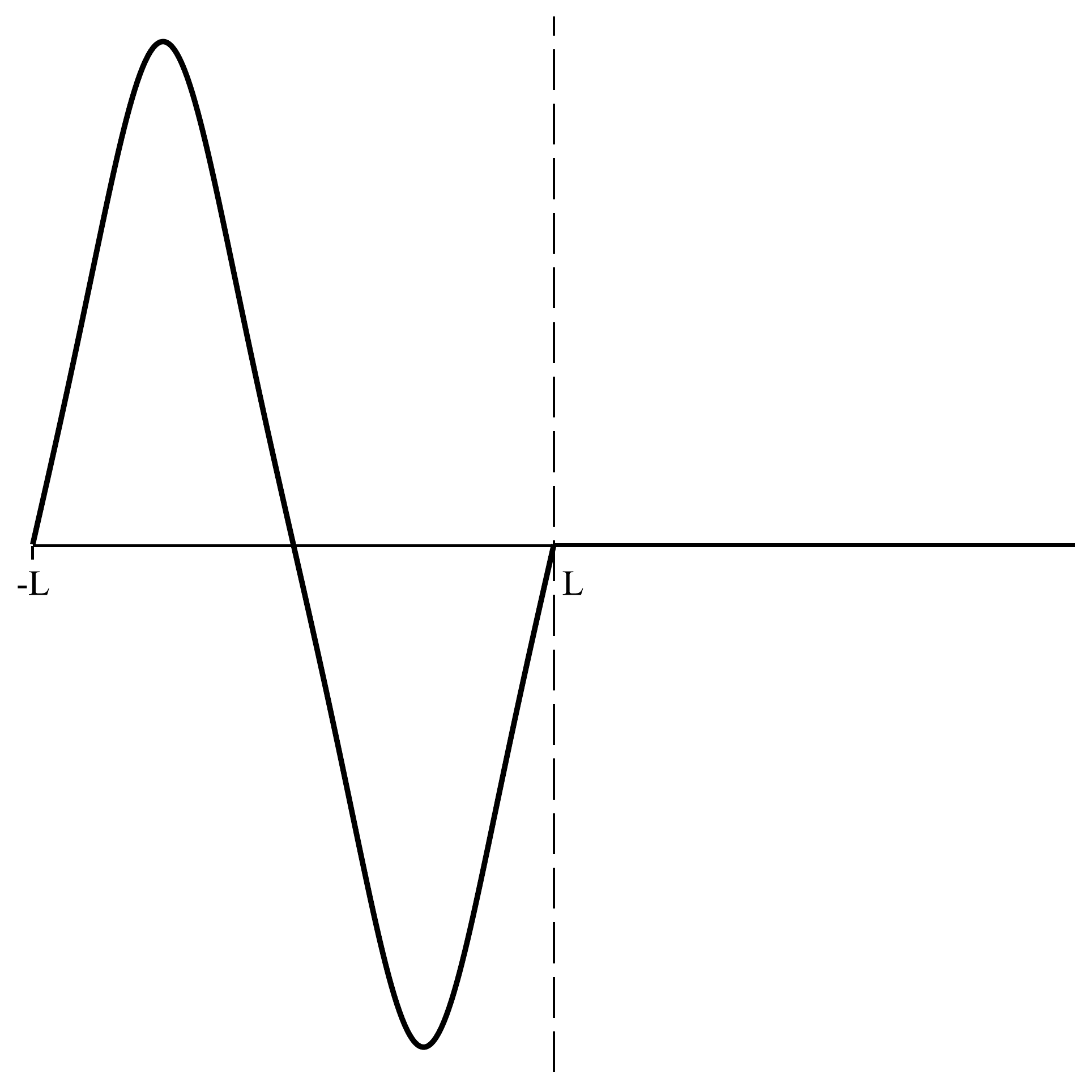}}
\fbox{\includegraphics[width=0.25\textwidth]{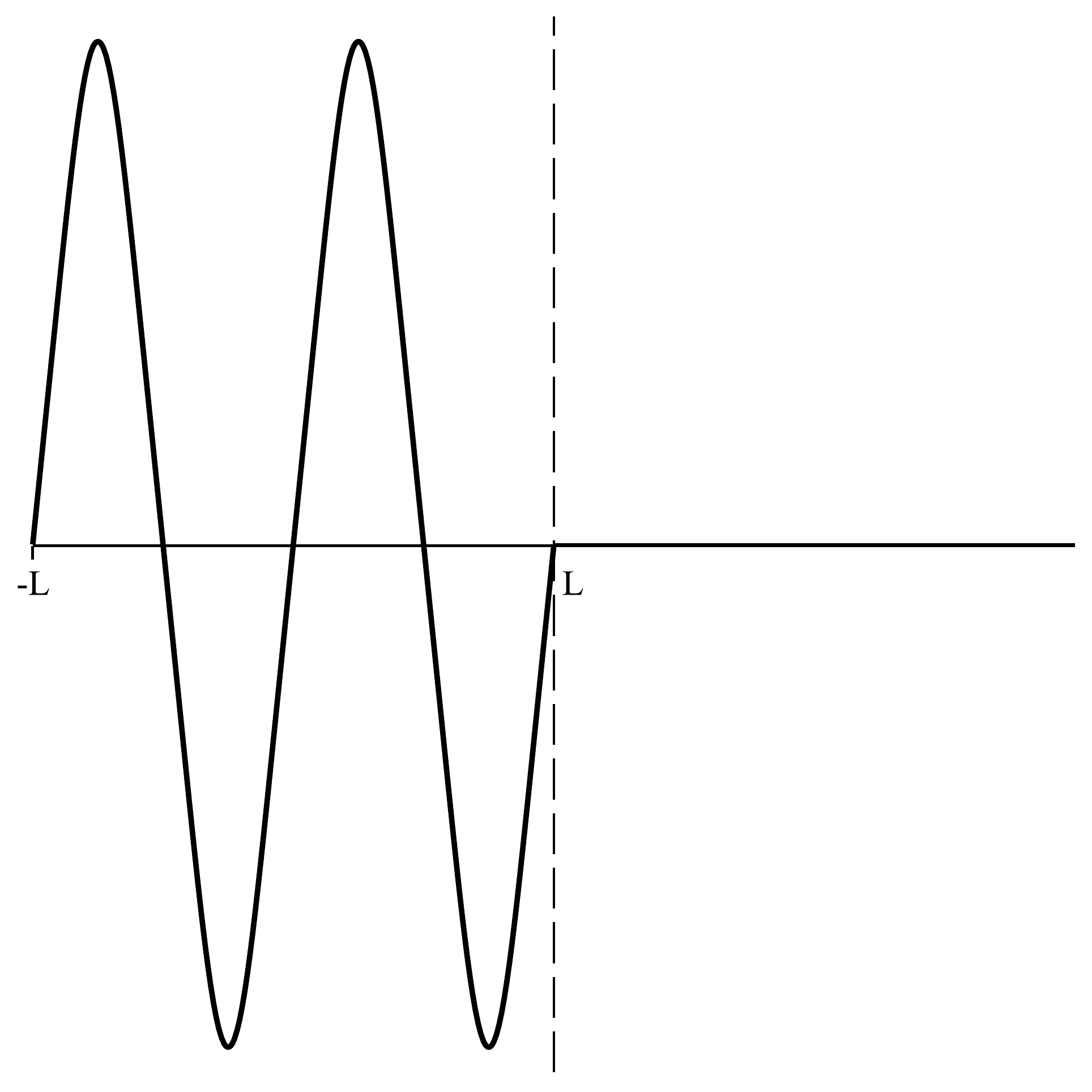}}\\
\vspace{4pt}
\fbox{\includegraphics[width=0.25\textwidth]{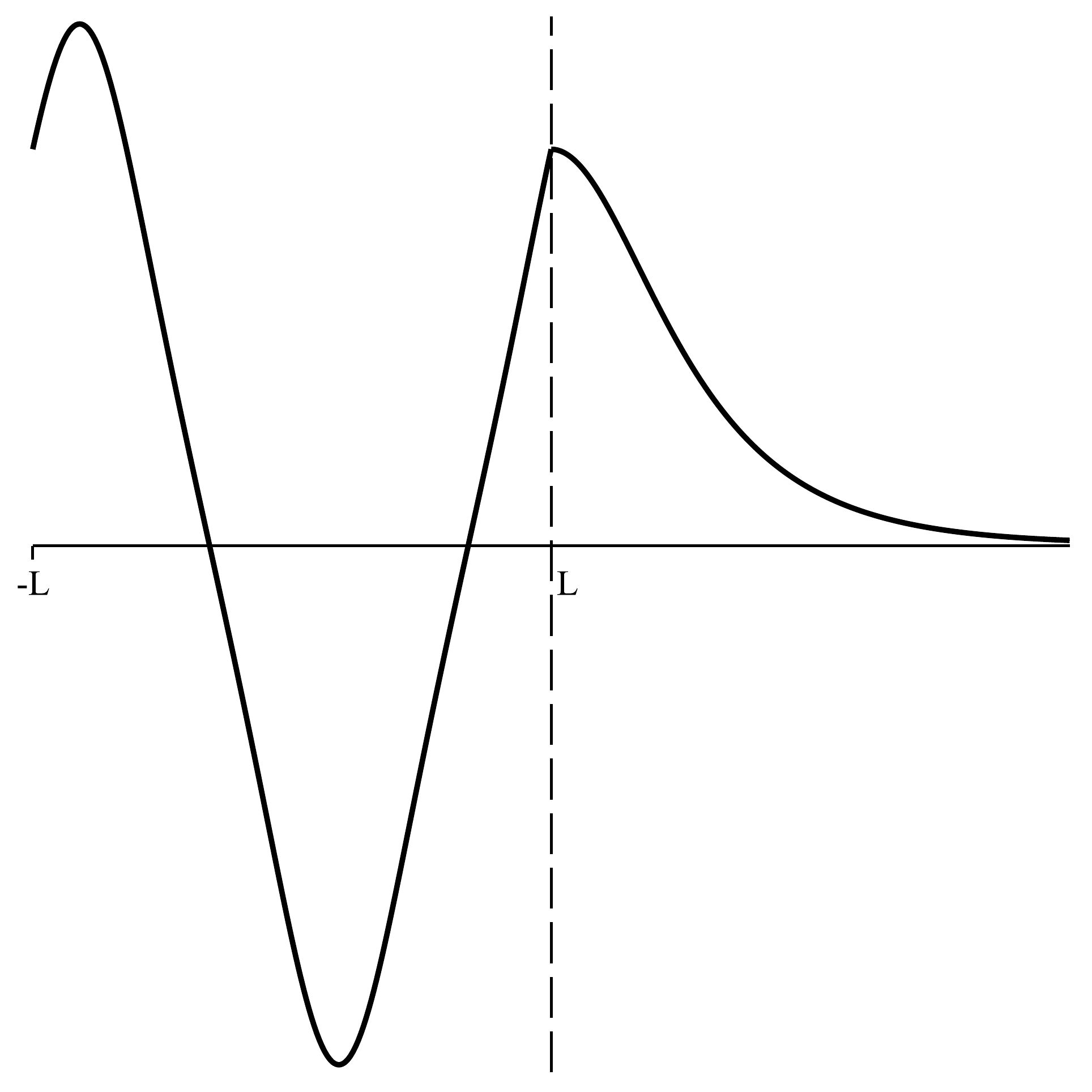}}
\fbox{\includegraphics[width=0.25\textwidth]{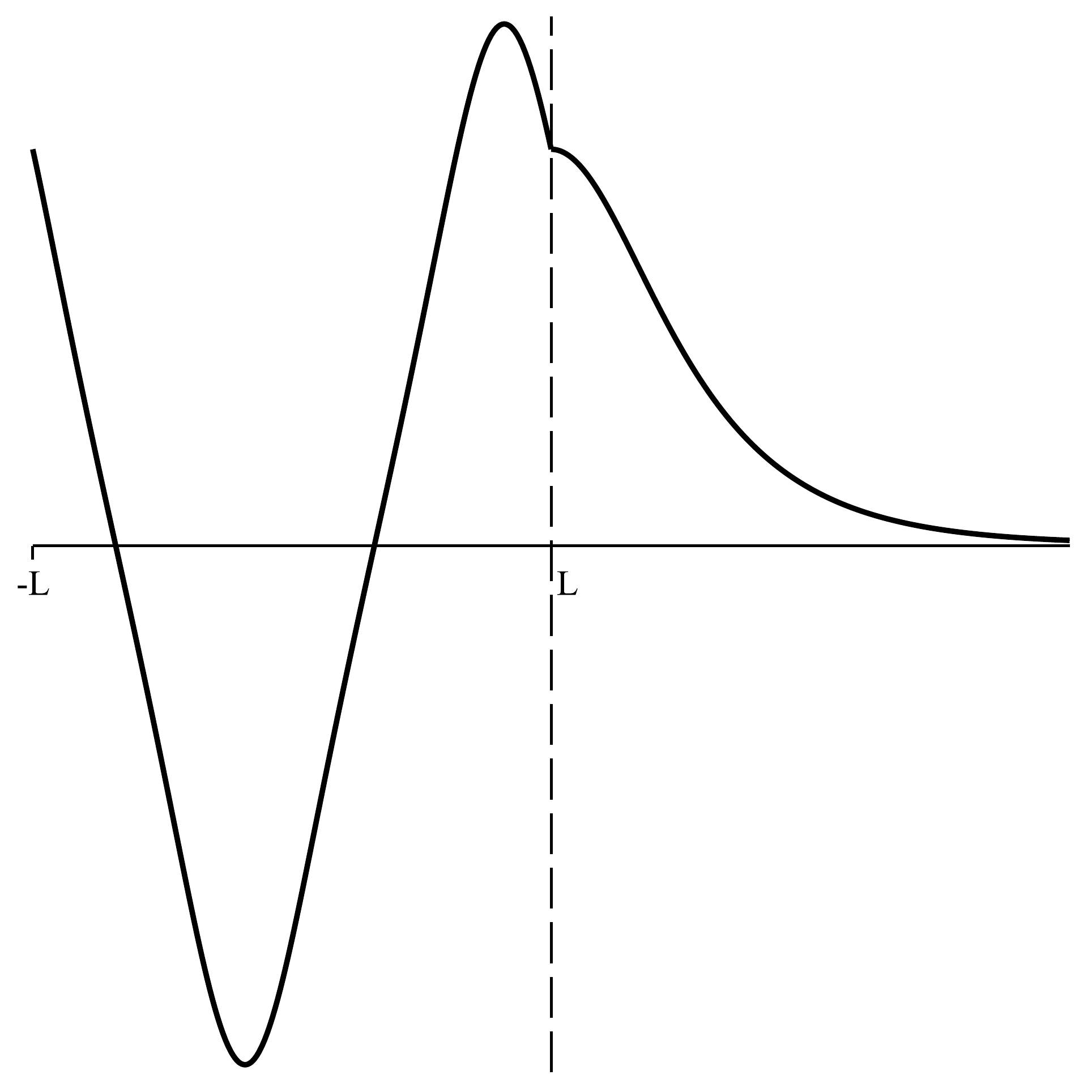}}
\caption{ \label{fig2}Top: plots of  representatives of cn-state $\Phi_{\ome,1}$ (left) and $\Phi_{\ome,2}$ (right). Bottom: plots of representatives of  cn-state  $\Phi_{\ome,1}^+$ (left)  and $\Phi_{\ome,1}^-$ (right).}
\end{figure}

Now we fix the free parameters.  A first family of  $\cn$-stationary states denoted by $\Phi_{\ome,n}$ vanishes on the tail. Notice that this is the only possible choice for $\omega>0$ because with this limitation the NLS on the tail has no square integrable solutions. We can choose $k$ such that the total length of the head is a multiple integer of
the period of $u_{cn}$, see Eq. \eqref{kn} below. Translating  $u_{cn}$ of $T_{cn}/4$ sets one of its nodes in the junction. Then, taking $\eta=0$, the Kirchhoff
b.c. are satisfied. The equation
\begin{equation}\label{kn}
2L = n\, T_{cn}( k) \qquad n\in \NA
\end{equation}
fixes $k_n(\ome)$ and $\Phi_{\ome,n}$ is  defined by
\[
\Phi_{\ome,n}(x,y) =\big( (-1)^n u_{cn}(x-L/2n ;k_n),0\big) \qquad \omega \in \RE,
\]
where the factor $(-1)^n$ is just a global phase factor which is there to match the notation used later on. See  Fig. \ref{fig2}, top part, for a qualitative plot of the first two eigenstates. 
Each plot should be understood in the following way:
on the left of the dashed line we draw the plot of $\Phi_{\ome}$ on the head of the
graph where $x$ runs from $-L$ to $L$; on the right we draw the plot of $\Phi_{\ome}$ on the tail of the graph, with the origin $y=0$ identified with $L$.
Notice the residual parity symmetry of the states $\Phi_{\ome,n}$:  the head-component is an odd function w.r.t. to the point $x=0$. The analysis of the behavior in $k$ of the function $2L/T_{cn}$ (it is enough to understand its monotonicity and range) leads to the following facts: Eq. \eqref{kn} admits solutions for any $n\in \NA$ if $\ome<0$; on the contrary, if $\ome>0$
it admits solutions only for $n$ such that  $\ome<\la_n$. Moreover we note that for $\ome < 0$ we have $k_n
^2\in(1/2,1]$ while for $\ome>0$ we have $k_n^2\in(0,1/2)$.  The constraint $\ome<\la_n$ when $\ome>0$ is related to the fact that 
$\Phi_{\ome,n}$ bifurcate from the vanishing state in correspondence of the embedded eigenvalues of the linear Hamiltonian $H$.
If we set $\ome = \la_n - \ve$ then as $\ve \to 0$, using known expansion of \eqref{K} near $k=0$ in  \eqref{kn}, we find the asymptotics $k_n=\sqrt{2\ve/(3\lambda_n)}+O(\ve^{3/2})$ and finally
\[
\Phi_{\ome,n}
=
c  \sqrt{\ve}  (-1)^n \Upsilon_n +O(\ve^{3/2}).
\]
Although we have embedded eigenvalues, we recover the typical behavior of a bifurcation into a gap:
the state $\Phi_{\ome,n}$ bifurcates, as $\ome$ crosses (in the decreasing direction) the threshold value $\la_n$,  from the empty state along the direction of the  eigenvector $\Upsilon_n$ corresponding to the embedded eigenvalue $\la_n$ of the linear operator $H$.

Next we discuss the case $\omega=0$. We note that  a  power expansion in the necessary existence condition \eqref{kn} gives
\[
k_n(\ome)^2 = \frac{1}{2} - \frac{L^2 \ome}{8n^2 K(1/\sqrt2)^2} + O(\ome^2).
\]
In this sense we have that for $\ome=0$, $k_n= 1/\sqrt{2}$ for all $n\in\NA$, and the solution  can be obtained by setting  $\frac{\omega}{1-2k^2} =   \frac {4n^2 K(1/\sqrt2)^2} {L^2}$ in the formula for $u_{cn}$. This gives 
\begin{equation}\label{ucn0}
u_{cn,0} (x) = \frac{2nK(1/\sqrt2)}{L}\cn\left(\frac{2nK(1/\sqrt2)}{L}x,\frac{1}{\sqrt2}\right).
\end{equation}
Then we have that for $\ome =0$ there exists an infinite family of stationary solutions labeled by $n\in\NA$ and they are explicitly given by 
 \begin{equation}\label{sol0}
 \Phi_{0,n} (x,y)= \left( (-1)^n u_{cn,0}\left(x-\frac{L}{2n}\right),0\right).
 \end{equation}
We conclude by noticing that  the family $\Phi_{\ome,n}$ contains an infinite number of stationary states for any $ \ome \in \RE$.

For $\ome<0$ there can be other solutions of the problem with non-trivial behavior on the tail, necessarily of the form $(u_{cn}(x-a;k),\phi(y-b))$, where  $a\in[0,T_{cn})$, $b\in\RE$  and $k\in(1/\sqrt2,1)$ are free parameters that must be adjusted to satisfy the vertex conditions \eqref{bc1} and \eqref{bc2}. In fact two further families $\Phi_{\ome,n}^+$ and  $\Phi_{\ome,n}^-$ with non-trivial behavior on the tail exist. After a straightforward analysis of the matching conditions one finds out that $b$ must equal zero and that $k$ must satisfy Eq. \eqref{kn}. Moreover,  shifting the argument in $u_{cn}$  to adjust the phase, one obtains the  explicit solutions
\[ 
\begin{aligned}
 \Phi_{\ome,n}^{+}(x,y) &=\big( (-1)^nu_{cn}(x-a_n;k_n),  \phi(y) \big) \\
 \Phi_{\ome,n}^{-}(x,y) &= \big( (-1)^n u_{cn}(x+a_n; k_n), \phi(y) \big)
\end{aligned} \qquad \omega < 0
\]
where $a_n$ is the smallest positive solution of 
\[
 \sqrt{\frac{k_n^2}{2k_n^2-1}}    \cn \left( \sqrt{\frac{|\ome|}{2k_n^2-1}} a_n;k_n \ri)=1.
\]
See  Fig. \ref{fig2}, bottom part, for a qualitative plot of the first eigenstate of each family.

These two families do not exist for $\ome >0$ since $\phi$ would not be defined.  The two families are the new branches of a pitchfork bifurcation from  $\Phi_{\ome,n}$ at the threshold of the continuum spectrum $\ome=0$. As $\ome \to 0^-$, we have 
$
a_n  = \frac{L}{2n} + O(\ome)
$
and 
\[
\Phi_{\ome,n}^\pm = \pm\Phi_{0,n} +O(\omega),
\]
where $\Phi_{0,n}$ are given in \eqref{sol0}.
Notice that the bifurcation is symmetry breaking because due to the presence of the tail and the consequent shift, the head-component of $\Phi_{\ome,n}^\pm$ lost the above recalled parity symmetry; notice moreover that both $\Phi_{\ome,n}$ and $\Phi_{\ome,n}^{\pm}$ have exactly $2n$ nodes. There are no other $\cn$-stationary states, in particular there are no $\cn$-stationary states when \eqref{kn} is not satisfied. \\

Now we discuss $\dn$-stationary states. In this case the tail-component has to be non vanishing since $\dn$ functions are separated from zero. So that solutions have the form
 $(u_{dn}(x-a;\kappa),\phi(y-b))$, where  $a\in[0,T_{dn})$,  $b\in\RE$, and $\kappa\in[0,1]$ are free parameters that again have to be fixed to satisfy  the boundary conditions \eqref{bc1} and \eqref{bc2}.
 
\begin{figure}[h!] 
\fbox{\includegraphics[width=0.25\textwidth]{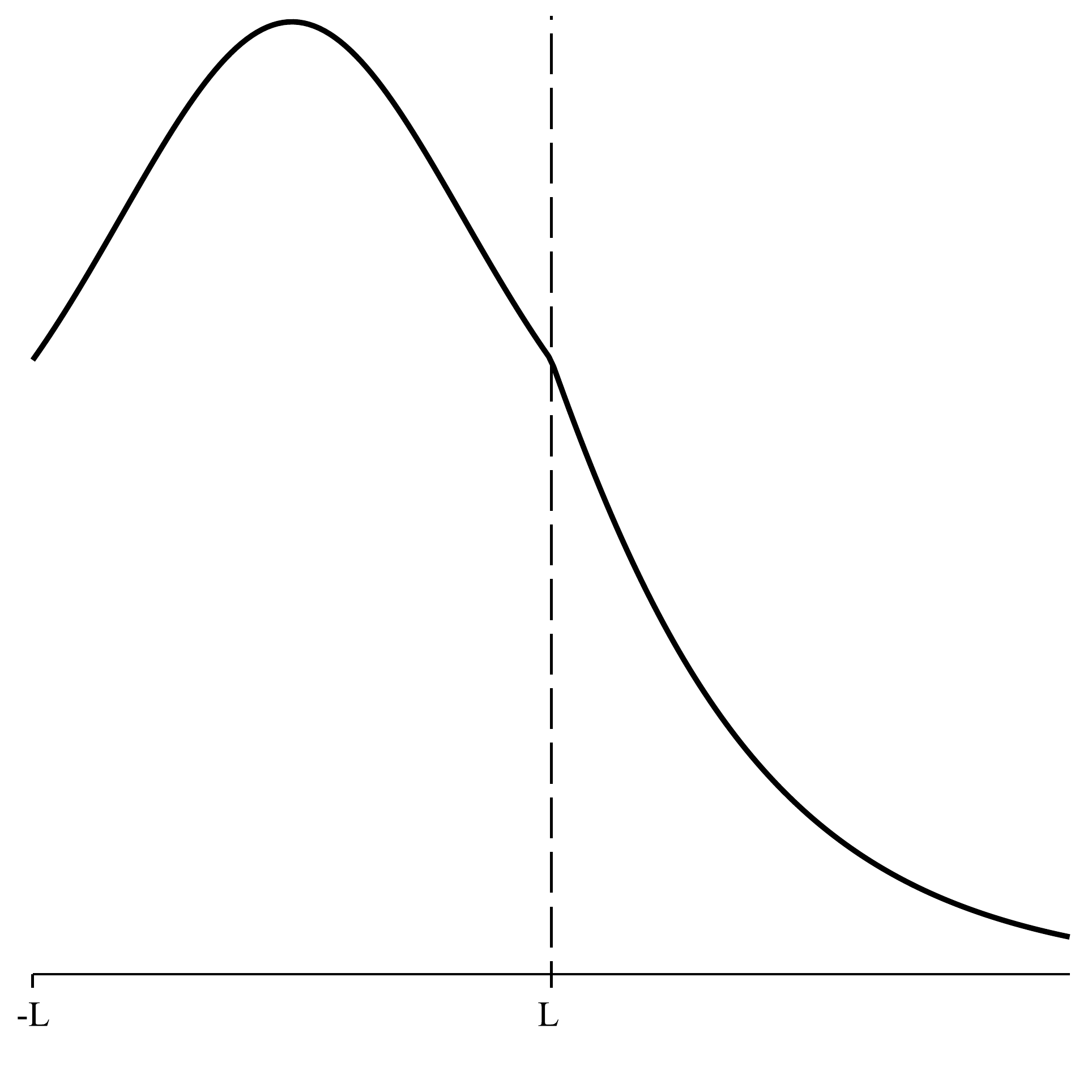}}
\fbox{\includegraphics[width=0.25\textwidth]{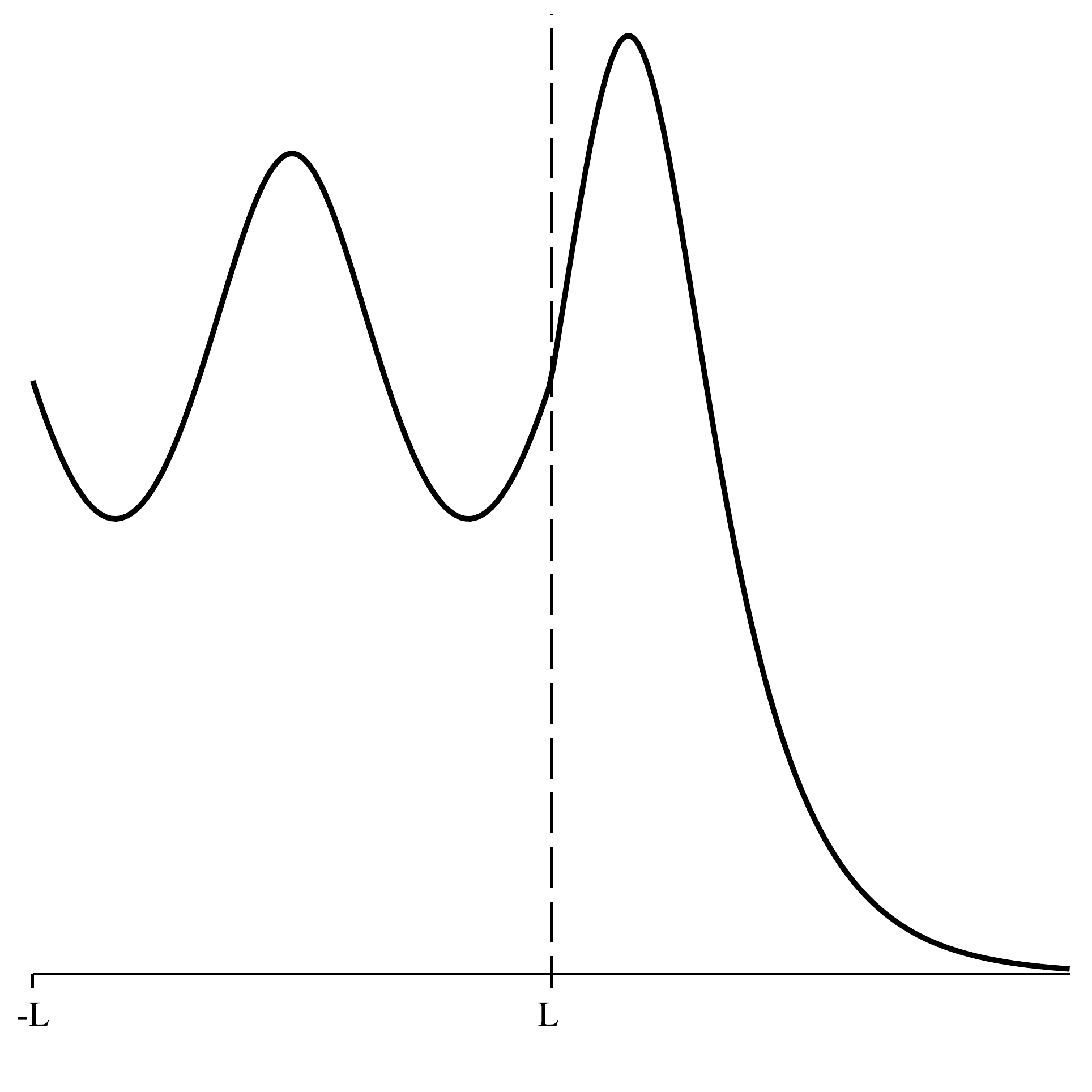}}\\
\vspace{4pt}
\fbox{\includegraphics[width=0.25\textwidth]{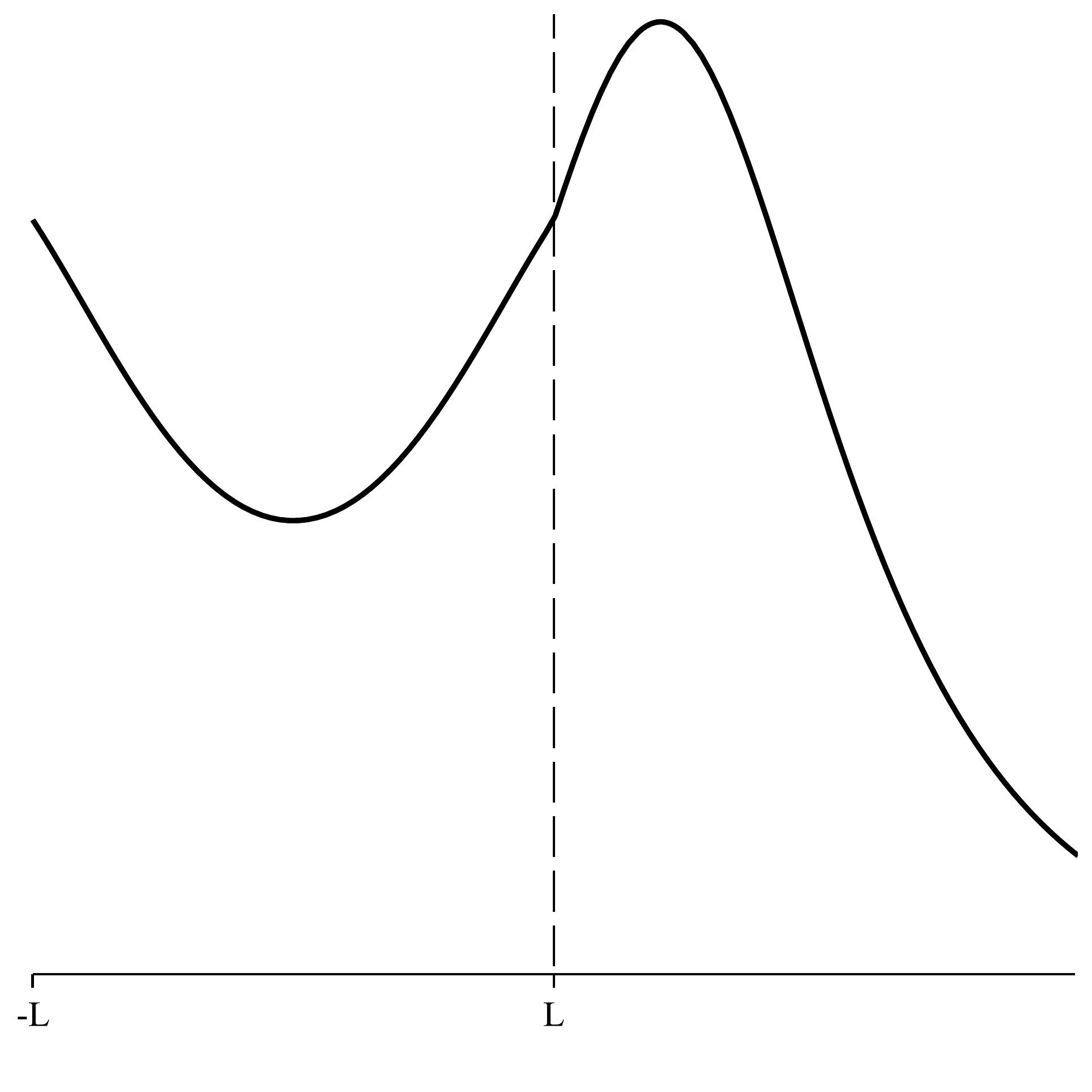}}
\fbox{\includegraphics[width=0.25\textwidth]{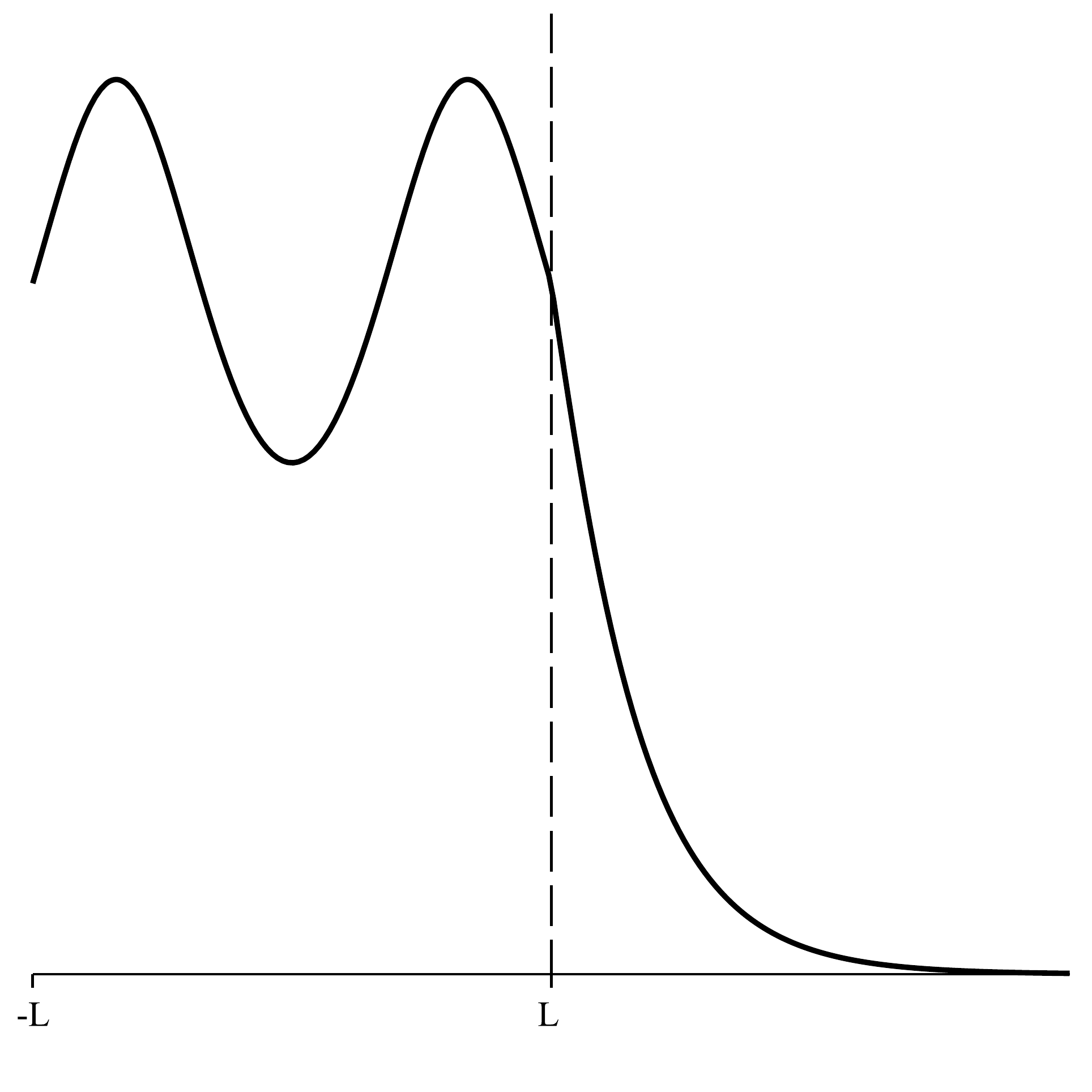}}
\caption{\label{fig3}Top: plots of representatives of dn-state $\Xi_{\ome,0,1}$ (left) and $\Xi_{\ome,0,2}$ (right). Bottom: plots of representatives of dn-state $\Xi_{\ome,1,1}$ (left) and $\Xi_{\ome,1,2}$ (right).}
\end{figure}

Imposing these matching conditions shows, after some lengthy analysis, that the parameter $a$ can assume only two values $a=0$ and $a=T_{dn}/2$. For this reason we distinguish two  families of $\dn$-solutions, both have non-vanishing component on the tail of the TG when $\ome<0$.
The first family has the following form
\begin{equation}\label{xizero}
\Xi_{\ome,0,n}(x,y)= \big( u_{dn}(x;\kappa_{0,n}),\phi(y- b_{0,n}) \big)  \qquad \omega <0 
\end{equation} 
for $n=1,2,\ldots$, and  where $\kappa_{0,n}$ are solutions (arranged in decreasing order) of the following equation
\begin{equation} \label{eqkn0}
\begin{split}
\frac{3 \kappa^4}{1-\kappa^2} &\cn^2 \lf( \frac{L\sqrt{|\ome|} }{\sqrt{2-\kappa^2}} ; \kappa \ri) \\ 
&\times
\lf( 1- \cn^2 \lf( \frac{L\sqrt{|\ome|} }{\sqrt{2-\kappa^2}} ; \kappa \ri) \ri)=1
\end{split}
\end{equation}
and $b_{0,n}$ is determined, up to a sign, by the equation 
\begin{equation}\label{shiftbzero}
\cosh^{-2} (\sqrt{|\ome|} b_{0,n} ) =  u_{dn}^2(L; \kappa_{0,n})/(2|\ome|).
\end{equation}
The sign is chosen as the same of $ u_{dn}' (L; \kappa_{0,n})$. The first $b_{0,n}$ is positive, the following couple is negative, the next couple is positive and so on. Notice that the sign of  $b_{0,n}$
determines if on the tail there is a decreasing function (negative  $b_{0,n}$) or a function with a maximum (positive  $b_{0,n}$). 
See Fig.  \ref{fig3}, top part, for a qualitative plot of the first two eigenstates of the first family.
Eq.  \eqref{eqkn0} has a finite number of solutions for any  $L\sqrt{|\ome|}$  and so is the number  of solutions $\Xi_{\ome,0,n}$. Moreover it always admits at least one solution and for 
$L\sqrt{|\ome|}\ll 1$ there is exactly one solution. On the contrary for $L\sqrt{|\ome|}\to \infty$ the number of solutions is divergent and growing as $L\sqrt{|\ome|}$.
The second family reads:
\begin{equation}\label{xiuno}\begin{split}
&\Xi_{\ome,1,n}(x,y) \\ 
&= \big(u_{dn}(x-T_{dn}(\kappa_{1,n})/2;\kappa_{1,n})   ,\phi(y- b_{1,n})\big),  
\end{split}\end{equation} 
with $\omega<0$,  $n=1,2,\ldots$, and where $\kappa_{1,n}$ are solutions (arranged in decreasing order) of the following equation
\begin{equation} \label{eqkn1}
\frac{3 \kappa^4 \cn^2 \lf( \frac{L\sqrt{|\ome|} }{\sqrt{2-\kappa^2}} ; \kappa \ri)
\lf( 1- \cn^2 \lf( \frac{L\sqrt{|\ome|} }{\sqrt{2-\kappa^2}} ; \kappa \ri) \ri) 
}{  \dn^{4} \lf( \frac{L\sqrt{|\ome|} }{\sqrt{2-\kappa^2}} ; \kappa \ri)  }
=1
\end{equation}
and $b_{1,n}$ is determined, up to a sign, by the equation
\begin{equation}
\label{shiftbuno}
\cosh^{-2} (\sqrt{|\ome|} b_{1,n} ) =\frac{  u_{dn}^2(L -T_{dn}(\kappa_{1,n})/2; \kappa_{1,n})}{2|\ome|}.
\end{equation}
The sign is chosen as the same of $ u_{dn}' (L-T_{dn}(\kappa_{1,n})/2; \kappa_{1,n})$, the first $b_{1,n}$ is negative, the following couple is positive, the next couple is negative and so on (opposite to $b_{0,n}$). 
See Fig. \ref{fig3}, bottom part, for a qualitative plot of the first two eigenstates of the second family.
In both families, the number of oscillations of the head-component increases with growing $n$.
Also Eq. \eqref{eqkn0} has a finite number of solutions for any 
$L\sqrt{|\ome|}$ 
and so is the number 
of solutions $\Xi_{\ome,1,n}$. For $L\sqrt{|\ome|}\to \infty$ it can be shown that the number of solutions diverges and its number grows as $L\sqrt{|\ome|}$.
On the contrary for $L\sqrt{|\ome|}\ll 1$ there are no solutions. Notice that Fig.  \ref{fig3} has to be understood only on a qualitative level: in general
$\kappa_{0,n} \neq \kappa_{1,n}$ and $b_{0,n} \neq b_{1,n}$.

Let us briefly consider the asymptotics of the  first dn-state $\Xi_{\ome,0,1}$ as $\ome$ approaches zero. Using the expansion $\cn(z,\kappa) = 1+O(z^2)$  in Eq. \eqref{eqkn0},  for $\ome$ small we get $\kappa_{0,1} = 1 + O(|\omega|)$. Then, using  the implicit definition \eqref{shiftbzero} of $b_{0,n}$, one can give for small $\omega$ the asymptotics of the shift $\sqrt{|\omega|} b_{0,1}$ of the tail:
\[
\sqrt{|\omega|}b_{0,1} = \sqrt{|\omega|}L + O(|\ome|^{\frac{3}{2}}).
\]
The expansions for $\kappa_{0,1}$ and $b_{0,1}$, together with the explicit definition \eqref{xizero}, give
\[
 \Xi_{\ome,0,1} = \sqrt{2|\ome|} (1,1) + O(|\ome|^{3/2}).
\]
Then we conclude that the state  $\Xi_{\ome,0,1}$ bifurcates from the vanishing state along the direction of the zero energy resonance $\Upsilon_{res}$ of the linear operator $H$. In this sense it is the only  dn-state with linear analogue. 

\section{Further remarks and conclusions}
\begin{figure}[h!]
\centering
\includegraphics[width=0.45\textwidth]{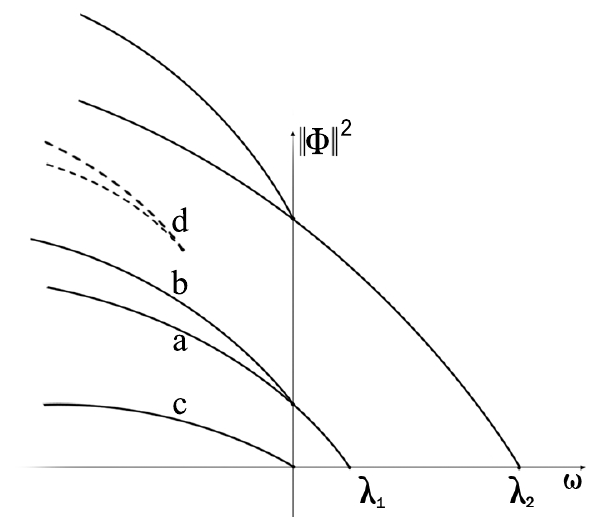}
\caption{\label{fig4}Bifurcation diagram}
\end{figure}

We summarize the main features of the aforementioned analysis with the bifurcation diagram in Fig. \ref{fig4}. 
The branch a) represents the embedded solitons bifurcating at the eigenvalues $\la_n$. The  branch b) represents the pitchfork bifurcation of $\Phi^\pm_{\ome}$
 at the threshold of the continuum spectrum. The branch c) represents the edge soliton $\Xi_{\ome,0,1}$
bifurcating at the threshold of the continuum spectrum. The dashed branches d) stand for a couple of higher modes of dn-eigenstates.
Although we have little information  about this last family of standing waves, from the analysis of Eqs. \eqref{eqkn0} and \eqref{eqkn1}, it turns out that, excluding the first one $\Xi_{\ome,1,1}$, they are generated in pairs in
the following sense. There is a first threshold $\al^\ast_1$, such that 
 $\Xi_{\ome,0,2}$ and  $\Xi_{\ome,0,3}$ exists for $L \sqrt{|\ome|}>\al^\ast_1$.
Moreover  $\Xi_{\ome,0,2}-\Xi_{\ome,0,3} \to 0 $ as  $L \sqrt{|\ome|}\to\al^\ast_1$
but  $\|\Xi_{\ome,0,2}\| $ and $\|\Xi_{\ome,0,3} \|$ do not converge to 0. The next couple
of states exhibits the same behavior with a different threshold $\al_2^\ast$ and so on.
This is true also  for the second  family,  for the states $\Xi_{\ome,1,n}$ with $n\geq 2$,  with
a second set of thresholds $\beta^\ast_k$.
Notice that the picture above is indicative of the bifurcations only, not of the behavior of $\frac{d}{d\omega}||\Phi||^2$.\\
We now discuss some perspectives and future developments which can be stimulated by the above analysis.

\subsection{Ground state and stability}
In the present paper we do not tackle the problem of the linear or orbital stability of the standing waves and these subjects will be discussed elsewhere. We only notice a related fact, as regard the variational properties of the standing waves. 
Let $M[\Psi] = \|\Psi\|^2$ be the mass  (or power, in nonlinear optics terminology) and 
\[\begin{split}
E[\Psi] =& \frac12 \int_{-L}^{L} |u'|^2 dx  - \frac14 \int_{-L}^{L} |u|^4 dx \\ 
& + \frac12 \int_{0}^{\infty} |\eta'|^2 dy  - \frac14 \int_{0}^{\infty} |\eta|^4 dy 
\end{split}\]
be the total energy associated to the state $\Psi = (u,\eta)$.  Define the ground state of the system as the minimizer, if it exists, of  the energy $E$ at fixed mass $M$.  

It has been shown recently (see \cite{AST}) that in a large class of networks with Kirchhoff boundary conditions at the vertices and sustaining a nonlinear Schr\"odinger dynamics a ground state does not exist. The TG does not belong to the class of graphs without ground state. We conjecture that for the TG the ground state exists and coincides with the first dn-state $\Xi_{\ome,0,1}$. If this were true a standard and well known consequence of energy minimization at constant mass would be the orbital stability of the state.

\subsection{Bifurcation conditions and resonances}
Another issue arisen by the previous analysis is the condition under which bifurcations occur, which again seems to display unconventional features. Let $\Sigma$ be any solution of the stationary equation \eqref{eq} (included the vanishing solution), and write $\Psi = \Sigma + X + i Y$ where $X$ and $Y$ are two real functions in the Hilbert space $L^2$. The linearization of the equation \eqref{eq} in a neighborhood of $\Sigma$ is described by the operator
$
L_\ome(\Sigma)( X + i Y)=L_{1,\ome}(\Sigma) X + i L_{2,\ome}(\Sigma) Y
$
with $L_{1,\ome}(\Sigma) = H - \ome - 3 |\Sigma|^2 $
and $
L_{2,\ome}(\Sigma) = H - \ome -  |\Sigma|^2 
$ with same boundary conditions as $H$. Here the ``potential'' $|\Sigma|^2$ must be understood as follows: let $\Sigma = (u,\eta)$ and $X=(v,\rho)$, then $|\Sigma|^2 X= (|u|^2 v, |\eta|^2 \rho)$. We note that whenever $\Sigma$ is a stationary state one has that zero is an eigenvalue of  $L_{2,\ome}$ and $\Sigma$ is the corresponding eigenvector. 
A commonly stated necessary condition for a bifurcation  is that $L_{1,\ome}(\Sigma)$ has an eigenvector with zero eigenvalue \cite{Y2}. This is certainly correct for the NLS on the whole space, but has to be completed and generalized to encompass examples such as the present one. The generalization should include the role of threshold resonances and not only proper eigenvectors of the linearization. 

Consider for example the above pitchfork bifurcation.  For $n\in\NA$, the operator $L_{1,0}(\Psi_{0,n})$ has no zero eigenvalues, but a zero energy resonance $\mathcal{X}_n^{res}$   given by 
\[
\mathcal{X}_n^{res}(x,y) = \left((-1)^n u_{cn,0}'\left(x-\frac{L}{2n}\right), \frac{(2nK(1/\sqrt2))^2}{L^2}  \right),
\]
where $u_{cn,0}$ was defined in Eq. \eqref{ucn0}.  Analogously, considering the lower dn-state $\Xi_{\ome,0,1}$ one easily recognizes for the linearization that $L_{1,0}(0) = L_{2,0}(0) =H$, the original linear Schr\"odinger operator on the tadpole. So both $L_{1,0}(0)$ and $L_{2,0}(0)$ have the same zero energy resonance as $H$, and no kernel.
Therefore both in the pitchfork bifurcation and the bifurcation of edge solitons the kernel of the linearization is not the expected one; instead, in the place of true eigenvectors appear zero energy resonances, which seem to represent an alternative signature of an ongoing bifurcation.

\subsection{Tadpole in a magnetic field}
Now we consider a tadpole in an external constant magnetic field $B$ which we take orthogonal to the plane of the ring and acting only on the ring itself. In this case the flux is constant and given by $\varphi=B L^2/\pi$ and the component of the vector potential $A$ tangential to the ring is given by $A=\frac{\varphi}{2L}$. The stationary NLS with a magnetic field for a state $\Psi=(v,\eta)$ now becomes
\begin{equation}\label{magneticeq}
\left\{\begin{aligned}
&\left(-i\frac{d}{dx} + A\right)^2v -|v|^2v = \omega v \quad &&x\in[-L,L]\\
-&\frac{d^2}{dy^2}\eta -|\eta|^2\eta = \omega \eta &&y\in[0,\infty)\\ 
&v(L)=v(-L)=\eta(0) &&\\ 
-&v'(L) + v'(-L) +\eta'(0)=0 &&
\end{aligned}\right. 
\end{equation}
with $\omega\in\RE$. Note that $\Psi$ satisfies the same boundary conditions at the vertex as before, which again assure selfadjointness or equivalently conservation of flux of particles.
As in the linear case (see \cite{E}), an elementary calculation shows that posing $v(x)=e^{-iAx}u(x)$ one has that the system  \eqref{magneticeq}  is equivalent to
\begin{equation}\label{sistema2}
\left\{\begin{aligned}
-&u''-|u|^2u = \omega u \quad &&x\in[-L,L]\\
-&\eta'' -|\eta|^2\eta = \omega \eta &&y\in[0,\infty)\\ 
&e^{-i\varphi/2}u(L)=e^{i\varphi/2} u(-L)=\eta(0)&&\\ 
-&e^{-i\varphi/2}u'(L) + e^{i\varphi/2}u'(-L) +\eta'(0)=0
&&
\end{aligned}\right. 
\end{equation}
with $\omega\in\RE$, and where we used $AL =\varphi/2$.  If the flux quantization condition
$\varphi=2n\pi$, $n\in\NA$, holds true, the boundary conditions in
\eqref{sistema2} read $(-1)^n u(L)=(-1)^n u(-L)=\eta(0)$ and $-(-1)^n
u'(L) + (-1)^n u'(-L) +\eta'(0)=0$. As a consequence there is a
one-to-one correspondence between the set of standing waves for the
problem without magnetic field and the set of standing waves for the
problem with magnetic field. More precisely, if $\varphi=2n\pi$,
$(u,\eta)$ is a standing wave for $B=0$ if and only if $((-1)^n
e^{-iAx}u,\eta)$ is a standing wave for $B\neq 0$.

We note that, due to the phase factor in the boundary conditions, the functions $u$ and $\eta$ which solve system \eqref{sistema2} are not necessarily real. After some analysis
one can show that $\eta$  can be chosen real and that for  real $u$ solutions exist only under the condition $\varphi=2n\pi$ and they are the ones described above.

The
situation for genuinely complex solutions is more complicated and it
will be studied in a complete way elsewhere. We just notice that posing $u(x)=R(x)e^{iS(x)}$ one gets an equation for
the ``one particle density'' $R^2(x)$ solved by elliptic functions and
the phase $S$ satisfies $S'=\frac{K}{R^2}$ where $K$ is an integration
constant; moreover a set of boundary conditions hold for the one
particle density and phase. Among those, there  is a generalization of the
flux quantization condition, which involves now the phase, $S(L)-S(-L)=n\pi + \varphi$.
This is a necessary but not sufficient  condition  to have standing
waves.

So flux quantization or more generally phase quantization suggest
that the magnetic field acts as a control parameter whose variation
can induce sudden and discrete formation of nonlinear bound
states. This poses the interesting problem of establishing which kind
of solutions exist when one approaches the critical quantization
values $\varphi=2n\pi$ (or more generally $\varphi=n\pi$). It is to be
noted that in the linear case the eigenvalues $2n\pi/L$ with odd $n$
are turned into resonance curves parametrized by the flux $\varphi$
(see \cite{E}), and so one would like to know if something similar to
resonant states exists in the nonlinear regime. 

 Notice that the
above considerations apply to more general nonlinearities than cubic,
i.e., every $U(1)$ invariant nonlinearity of the kind
$N(u)=G(|u|^2)u$.

\subsection{Other couplings at junction} Finally, different coupling at the vertex could be
studied. For example one could consider some physical device which
produces a perturbation at the vertex of attractive or repulsive
kind. The easiest way to embody this effect is to modify the boundary
condition introducing a delta potential at the vertex, which is
modeled by the same continuity condition but has a different condition
on the derivative, i.e., $-u'(L)+u'(-L)+\eta'(0)=\alpha u(L)$, where
$\alpha$ is a real constant representing the  ``strength''  of the delta potential. It is known that
these boundary conditions result as a limit of a regular potential
concentrated at the vertex when its support shrinks at the vertex
itself.  As for the case with Kirchhoff boundary conditions, $ \la_n= (n\pi/L)^2$ are embedded eigenvalues with corresponding eigenfunctions given by $\Upsilon_n$, see Eq.  \eqref{Ups}. This is a direct consequence of the fact that the additional $\alpha$ term in the boundary conditions is ineffective when there is no tail.
Analogously, for the nonlinear problem, the cn-type solutions   $\Phi_{\omega,n}$, having  vanishing tail,  persist as well.  If the delta potential strength $\alpha$
is negative (attractive well in the junction), this generates a simple isolated negative
eigenvalue in the linear problem. By standard bifurcation theory (see
\cite{Pel}), this will give rise to a new curve of nonlinear bound
states related to the corresponding linear eigenvector. An open
problem is the existence and characterization of other bound states
and their relation with the linear problem.\\

 To conclude, the
classification of nonlinear bound states for the cubic NLS on the TG
exhibits a variety of behaviors previously unknown for standard NLS
with power nonlinearity on the line with or without
inhomogeneities. Moreover when an external magnetic field is
introduced, an analogous scenario displays, when a quantization
condition involving the magnetic flux holds true. This indicates the
possibility of an experimental induction and control of long living
states for this system. The analysis of NLS equation on the simplest
topologically non-trivial structure given by the tadpole graph could
suggest that analogous phenomena could be present and studied in more
complex networks.\\

\section*{Acknowledgments}
The authors benefitted from discussions with Pavel Exner, Dmitry Pelinovsky and Uzy Smilansky. 
D.F. and D.N.  acknowledge the support of the FIRB 2012 project, code RBFR12MXPO. C.C. acknowledges the support of the FIR 2013 project, code RBFR13WAET.

\end{document}